\documentclass[journal]{IEEEtran}
\usepackage{xcolor,soul,framed} 

\colorlet{shadecolor}{yellow}
\usepackage{graphicx}
\graphicspath{{../pdf/}{../jpeg/}}
\DeclareGraphicsExtensions{.pdf,.jpeg,.png}

\usepackage[cmex10]{amsmath}
\usepackage{array}
\usepackage{mdwmath}
\usepackage{mdwtab}
\usepackage{eqparbox}
\usepackage{url}
\usepackage[noadjust]{cite}
\usepackage[T1]{fontenc}
\usepackage{graphics} 
\usepackage{epsfig} 
\usepackage{mathptmx} 
\usepackage{times} 
\usepackage{amsmath} 
\usepackage{amssymb}  
\usepackage{multirow}
\usepackage{placeins}
\usepackage{tabularx}
\usepackage{booktabs}
\usepackage{bigstrut}
\usepackage{array}
\usepackage{eqnarray}
\usepackage{makecell}
\usepackage{epstopdf}
\usepackage{etoolbox,siunitx}
\robustify\bfseries
\usepackage[flushleft]{threeparttable}
\usepackage{romannum}

\AppendGraphicsExtensions{.tif}
\newcolumntype{P}[1]{>{\centering\arraybackslash}p{#1}}
\newcolumntype{L}[1]{>{\raggedright\let\newline\\\arraybackslash\hspace{0pt}}m{#1}}
\newcolumntype{C}[1]{>{\centering\let\newline\\\arraybackslash\hspace{0pt}}m{#1}}
\newcolumntype{R}[1]{>{\raggedleft\let\newline\\\arraybackslash\hspace{0pt}}m{#1}}
\newcolumntype{d}[1]{D{.}{.}{#1}}
\newcolumntype{s}[1]{S[table-format=2.2, table-column-width=#1]}

\newcounter{rCounter}
\newcommand{\resHeader}[1]{\refstepcounter{rCounter} \textit{\textbf{Result~\Roman{rCounter}: #1}} \par}

\usepackage{xr-hyper}
\makeatletter
\newcommand*{\addFileDependency}[1]{
  \typeout{(#1)}
  \@addtofilelist{#1}
  \IfFileExists{#1}{}{\typeout{No file #1.}}
}
\makeatother

\newcommand*{\myexternaldocument}[1]{%
    \externaldocument{#1}%
    \addFileDependency{#1.tex}%
    \addFileDependency{#1.aux}%
}
\usepackage[hidelinks]{hyperref}

\myexternaldocument{supplementary}

\begin{document}
\title{FBCNet: A Multi-view Convolutional Neural Network for Brain-Computer Interface}
  \author{Ravikiran Mane \IEEEmembership{Member,~IEEE,}
  Effie Chew, 
  Karen Chua, 
  Kai Keng Ang \IEEEmembership{Senior Member,~IEEE,} 
  Neethu Robinson \IEEEmembership{Member,~IEEE,}
  A. P. Vinod \IEEEmembership{Senior Member,~IEEE,}
  Seong-Whan Lee \IEEEmembership{Fellow,~IEEE,} and 
  Cuntai Guan \IEEEmembership{Fellow,~IEEE}

\thanks{Ravikiran Mane, Kai Keng Ang, Neethu Robinson, Cuntai Guan are with  the Nanyang Technological University, 50 Nanyang Avenue, Singapore ({ravikian001@e.ntu.edu.sg, kkang@ntu.edu.sg, nrobinson@ntu.edu.sg, ctguan@ntu.edu.sg}).}%
\thanks{Effie Chew is with National University Hospital, Singapore. ({effie\_chew@nuhs.edu.sg})}%
\thanks{Karen Chua is with Tan Tock Seng Hospital, Singapore. ({karen\_chua@ttsh.com.sg})}%
\thanks{Kai Keng Ang is with Institute for Infocomm Research, Agency for Science, Technology and Research (A*STAR), Singapore. ({kkang@i2r.a-star.edu.sg})}%
\thanks{A. P. Vinod is with Indian Institute of Technology, Palakkad, India. ({vinod@iitpkd.ac.in})}
\thanks{ S. -W. Lee is with the Department of Artificial Intelligence, Korea University, Seoul 02841, South Korea ({sw.lee@korea.ac.kr})}
\thanks{Corresponding author: Cuntai Guan}}  

\maketitle

\begin{abstract}
Lack of adequate training samples and noisy high-dimensional features are key challenges faced by Motor Imagery (MI) decoding algorithms for electroencephalogram (EEG) based Brain-Computer Interface (BCI). To address these challenges, inspired from neuro-physiological signatures of MI, this paper proposes a novel Filter-Bank Convolutional Network (FBCNet) for MI classification. FBCNet employs a multi-view data representation followed by spatial filtering to extract spectro-spatially discriminative features. This multistage approach enables efficient training of the network even when limited training data is available. More significantly, in FBCNet, we propose a novel Variance layer that effectively aggregates the EEG time-domain information. With this design, we compare FBCNet with state-of-the-art (SOTA) BCI algorithm on four MI datasets: The BCI competition IV dataset 2a (BCIC-IV-2a), the OpenBMI dataset, and two large datasets from chronic stroke patients. The results show that, by achieving 76.20\% 4-class classification accuracy, FBCNet sets a new SOTA for BCIC-IV-2a dataset. On the other three datasets, FBCNet yields up to 8\% higher binary classification accuracies. Additionally, using explainable AI techniques we present one of the first reports about the differences in discriminative EEG features between healthy subjects and stroke patients. Also, the FBCNet source code is available at \url{https://github.com/ravikiran-mane/FBCNet}.
\end{abstract}

\begin{IEEEkeywords}
Brain Computer Interface, Stroke Rehabilitation, Deep Learning, Motor Imagery Classification
\end{IEEEkeywords} 

\section{Introduction}
In recent years, the technology of Brain-Computer Interface (BCI) has emerged as a powerful means of communication and control for severely paralyzed patients \cite{Pfurtscheller2006, mane2020bci}. BCI systems capture real-time brain activity and try to decode users' intentions from the observed neuronal activations \cite{Pfurtscheller2006}. The decoded state is then used to control external devices \cite{Ang2012c, Ang2015} or for communication \cite{Li2010}. In BCI systems, electroencephalography (EEG) is the most widely used signal acquisition modality and Motor-Imagery (MI) based EEG-BCI, wherein participant performs mental rehearsal of a particular motor movement is one of the frequently investigated protocols. Furthermore, by BCI-monitored repetitive MI training, MI-BCI systems have also shown promising results in post-stroke motor rehabilitation \cite{Ang2015, mane2020bci}.  Therefore, owing to its high clinical relevance, EEG-BCI literature contains many reports on decoding techniques to classify various MI classes with high accuracy. Classical machine learning techniques like linear and non-linear classifiers, nearest-neighbour classifiers as well as more data-driven techniques like neural networks and deep learning have been explored for this task of MI classification \cite{Ang2008, Schirrmeister2017, Sakhavi2018, Lawhern2018, Kwon2019}.

Noisy and high dimensional data, high inter-trial variance, and scarcity of training data are among the key challenges faced by the EEG-BCI classifiers \cite{roa2013}. Furthermore, the intrinsic nature of MI and high intra-class variability in MI signature make the task of MI classification even more difficult. Owing to all these issues, the MI classification is considered to be among the most challenging tasks in the BCI domain \cite{Lotte2018, Roy2019}. The existing works on MI classification can be grouped into two approaches; the classical machine learning approaches, and deep learning based approaches.  

The classical machine learning approaches have commonly focused on the extraction of neuro-physiologically sound features from EEG data to achieve higher classification accuracies. These algorithms generally employ a multistage approach wherein the EEG data is first preprocessed to reduce the noise and then neurophysiologically sound features are extracted from the data to enhance SNR and reduce dimensionality. Neuroscientific studies have documented that MI elicits characteristic EEG activation patterns known as sensory-motor rhythms (SMR). SMRs are generally observed at the contralateral and ipsilateral sensory-motor regions in the form of a time-locked, event-related desynchronization/ synchronization (ERD/ERS) \cite{Pfurtscheller2006}. It is also known that different classes of MI differ in the spectro-spatial distribution of SMRs \cite{Pfurtscheller2006}. By extracting these features, algorithms like Filter Bank Common Spatial Patterns (FBCSP) \cite{Ang2008}, as well as Riemannian geometry based methods \cite{Zhang2020} have achieved high accuracies in this scenario of limited and noisy EEG training data. However, they still suffer from the high susceptibility to intra-trial variance and have a high dependence on handcrafted features. 

In recent years, deep learning, which is an extensively data-driven approach to classification has shown very promising results in the EEG classification domain. Deep learning methods offer unique advantage of end-to-end processing capability and they eliminate the need for handcrafted feature extraction. Therefore, deep learning architectures, particularly based on Convolutional Neural Network (CNN), have gained popularity in the BCI domain due to their ability of effectively learning the local connectivity patterns from the given data \cite{Schirrmeister2017, Sakhavi2018, Lawhern2018, Kwon2019}. These architectures have outperformed the classical machine learning techniques, but improvements achieved are still marginal \cite{Schirrmeister2017, Sakhavi2018, Lawhern2018, Kwon2019}. These marginal improvements can be attributed to the scarcity of training data and the high feature dimensionality of the MI signals which together result in heavy overfitting of deep learning models and create unique challenges for adaptation of these methods in the BCI field \cite{Roy2019}. Considering these advantages and limitations of deep learning and classical machine learning methods, hybrid methods may hold the potential to achieve the best classification accuracies in the MI-BCI field.  

Furthermore, although stroke patients are among the most-important target population, exploration of deep learning techniques for MI classification is still largely limited to the data from healthy people. Alteration of brain dynamics, changes in EEG brain rhythms, and modifications in motor function control are some of the known effects following stroke \cite{Boyd2017, mane2019, mane2018quantitative}. Furthermore, it has been documented that close to 13\% stroke patients can not control the MI-BCI using classical machine learning techniques \cite{Ang2011}. Therefore, it is necessary to investigate how deep learning architectures perform in the stroke population and if they can extend the technology of MI-BCI to more patients by achieving better classification accuracy. One initial report indicates that deep learning architectures may perform worse than classical machine learning techniques in stroke patients, but this result needs further verification with data from a large stroke population \cite{Raza2020}. 

Lastly, the use of deep learning in patient population also greatly necessitates interpretability of the classification decisions. In stroke patients, since the aim is to restore brain activation to a healthy state, it is important to evaluate the information encoded by the neural network. Therefore ease of decision explanation is another important parameter when it comes to patients' data.

Considering all the above challenges, in this paper, we propose Filter-Bank Convolutional Network (FBCNet), which is a novel end-to-end CNN architecture for subject-specific MI classification. In FBCNet architecture we take a hybrid approach; we leverage the recent advances in deep learning technologies and mitigate their shortcomings by encoding the neurophysiological priors of MI in the architecture design. FBCNet encodes spectro-spatial discriminative information associated with the MI with the help of spectral filtering of the EEG and CNN based spatial filtering. Next, we propose a novel Variance layer for effective extraction of distinct MI signatures encoded by the temporal fluctuations and dimensionality reduction, which results in extremely compact architecture. While being simple and interpretable, FBCNet also offers significantly higher classification results. The preliminary results of MI classification using this network are presented in \cite{Mane2020}. In this paper, we further investigate the classification superiority of FBCNet over other deep learning architectures and FBCSP with the help of two publicly available datasets of MI classification. The analysis is further extended to two MI datasets covering a total of 71 chronic stroke patients. In addition, using explainable AI techniques, we discuss the difference between the data from healthy and stroke patients in the context of MI classification. Following are the main contributions of this paper:
\begin{itemize}
    \item A compact and neurophysiologically inspired CNN architecture named FBCNet is proposed for MI classification.
    \item A novel Variance layer is proposed for effective extraction of EEG temporal information and parameter reduction. 
    \item We present one of the first reports on the comparison between classical machine learning algorithms and deep learning architectures for MI decoding in a large population of chronic-stroke patients.
    \item We present one of the first reports on the effectiveness of deep learning architectures for MI decoding in a large population of chronic-stroke patients.
    \item We show that, for stroke patients, classical machine learning approaches may outperform the general-purpose deep learning architectures and careful fusion of deep learning methods and neurophysiological knowledge of MI, as done in FBCNet, can achieve the best classification accuracies for both healthy subjects and stroke patients.  
\end{itemize}

The PyTorch implementation of the FBCNet is available at \url{https://github.com/ravikiran-mane/FBCNet}. 

\section{Related Works}
\begin{figure*}[tb]
      \centering
      \includegraphics[width=7in]{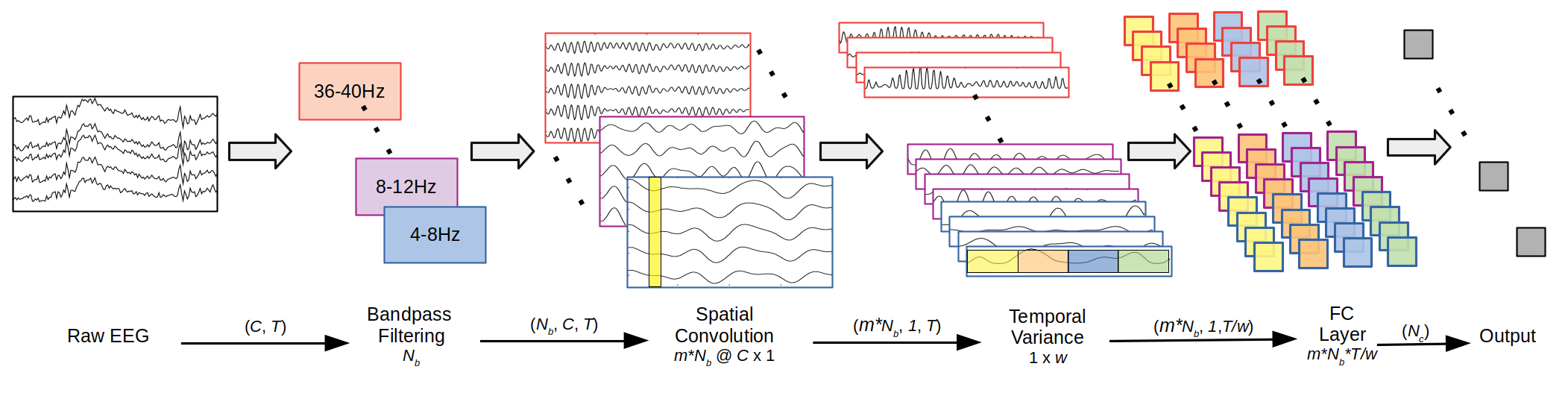}
      \caption{Proposed network architecture: FBCNet. ($C$: number of EEG channels, $T$: number of time points, $N_b$: number of frequency bands, $m$: number of convolution filters per frequency band, $N_c$: number of output classes) }
      \label{network}
\end{figure*}

Many classical machine learning techniques have been proposed for EEG-MI classification and an extensive summary of them can be found in \cite{Lotte2018}. Among them, FBCSP is one of the most successful algorithms \cite{Ang2008} and it shares a similar design philosophy as that of FBCNet. Therefore, we directly compare the results of FBCNet with FBCSP. 

In recent years, many deep learning architectures have also been proposed for use in the EEG-BCI domain \cite{Schirrmeister2017, Robinson2019, Lawhern2018, Kwon2019, Sakhavi2018}. Among them, Deep ConvNet \cite{Schirrmeister2017} and EEGNet \cite{Lawhern2018} are two architectures that have seen widespread adaptation in the EEG community and have provided open-source code implementations. Therefore, we use these two architectures to compare against the FBCNet in all the evaluation datasets. Lastly, we compare the performance of FBCNet against many recent architectures based on their reported accuracies on the BCI competition IV-2a dataset.  

\section{Methodology}

\subsection{Proposed Architecture : Filter-bank Convolutional Net}
\label{FBCNet:description}
FBCNet is designed with the aim of effectively extracting the spectro-spatial discriminative information, which is the signature of MI while avoiding the problem of overfitting in the presence of small datasets. At its core, FBCNet architecture is composed of the following four stages: 

\begin{enumerate}
    \item Multi-view data representation: The multi-view representation of the EEG data is obtained by spectrally filtering the raw EEG with multiple narrow-band filters. 
    \item Spatial transformation learning: The spatial discriminative patterns for every view are then learned using a Depthwise Convolution layer. 
    \item Temporal feature extraction: Following spatial transformation, a novel Variance layer is used to effectively extract the temporal information.
    \item Classification: A fully connected (FC) layer finally classifies features from the Variance layer into given classes.
\end{enumerate}

The multi-view EEG representation followed by the spatial filtering allows extraction of spectro-spatial discriminative features and the Variance layer provides a compact representation of the temporal information. With this brief design philosophy, this sub-section provides details of the FBCNet architecture which is presented in Fig. \ref{network} and summarized in supplementary material Table S1.   

\subsubsection{Spectral Localization by Multi-view Data Representation}
Consider a single-trial raw EEG data represented as $ x \in \mathbb{R}{^{C \times T}} $ and its corresponding label $ y \in \{0,1,..,N_c-1\}$, where $C$ represents number of EEG channels, $T$ represents time points and $N_c$ is total number of distinct classes.

It is known that MI related information in EEG data is spectrally localized with most information being present in the mu (8-12 Hz), and beta (12-32Hz) bands \cite{Pfurtscheller2006}. Therefore, to localize this discriminative information, in its first stage, FBCNet creates a multi-view representation of the broad-band EEG data wherein each view represents a narrow-band localized EEG. The multi-view representation, $  x_{FB} \in \mathbb{R}{^{N_b \times C \times T }} $, is generated by spectrally filtering the raw EEG $x$ with a filter bank $F = {\{f_i\}}_{i=1}^{N_b}$ consisting of $N_b$ number of narrow-band temporal filters. Following the filtering operation, the time-series along the third dimension of $ x_{FB}$ becomes spectrally localized.
Therefore, 
\begin{equation}
    x_{FB} = F \otimes x  \hspace{1em} \in \mathbb{R}{^{N_b \times C \times T }}
\end{equation}
Where, $\otimes$ indicates bandpass filtering operation. 

The filter bank $F$ can consist of any number of filters with varying cut-off frequencies. In this particular work, the filter bank is constructed using $N_b = 9$ filters with non-overlapping frequency bands, each of 4Hz bandwidth, spanning from 4 to 40 Hz (4-8, 8-12, ..., 36-40 Hz). The filtering is done using the Chebyshev Type II filter with a transition bandwidth of 2Hz and a stopband ripple of -30dB. This selection of filter bank was motivated by the traditional division of EEG in neurologically significant spectral bands that was proposed in the FBCSP algorithm and it has been shown to achieve good classification accuracies across multiple works \cite{Ang2008, Ang2012d, Ang2012c}. 

\subsubsection{Spatial Localization by CNN}

Following the deterministic spectral localization, spatial localization of EEG discriminative features is achieved by a Spatial Convolution Block (SCB) comprising of a Convolution layer, a Batch Normalization layer \cite{Ioffe2015} and a Swish nonlinearity \cite{swish}. FBCNet uses a Depthwise Convolution layer \cite{Chollet2016} of kernel size $ = (C, 1)$ for learning spatially discriminative patterns. Also, the use of Depthwise Convolutions results in each of these filters being associated with EEG from only one frequency band and, the depth parameter, $m$, controls the number of spatial filters per frequency band. Also, the Convolution kernel that is sized to span across all the channels effectively acts as a spatial filter. Following the CNN layer, we apply Batch-normalization along the feature map dimension before applying the Swish nonlinearity. We also regularize each convolutional kernel by using a maximum norm constraint of 2 on its weights; $||w||^2 < 2$ (weight-normalization). Effectively, SCB outputs $m \times N_b$ time-series $ x_{SCB} \in \mathbb{R}{^{(m \times N_b) \times 1 \times T}} $ which are spectro-spatially localized. In the default FBCNet structure, we set the value $m$ to be equal to be 32. We also evaluate the effect of different values of $m$ on model performance in ablation analysis. 

\subsubsection{Temporal Feature Extraction by Variance Layer}
The raw EEG data generally contains a large number of features along the time dimension and these features present the most amount of intra-class variance and high noise content. Therefore, to avoid overfitting of classification models, it is necessary to reduce the time dimensional features by effective extraction of the most relevant temporal information. Max or Average Pooling strategies are the most commonly employed techniques for this purpose of reduction in feature dimensions \cite{Schirrmeister2017, Robinson2019, Lawhern2018}. However, considering that various classes of MI differ in their spectral power (ERD/ERS), a variance operation, which represents the spectral power in the given time series becomes a more suitable option for EEG temporal characterization. Therefore, for effective extraction of temporally discriminative information, we propose a novel Variance layer that characterizes a time series by computing its variance. So, in the forward pass, for any time-varying signal, $g(t)$, the output of the Variance layer is given by,
\begin{equation}
 v = Var(g(t)) = \dfrac{1}{T} \sum_{t = 0}^{T-1} (g(t) - \mu)^2
\end{equation}
where, $T$ is total number of time-points and $\mu$ is the mean of $g(t)$.

The effect of the Variance layer is yet more significant for the EEG data during the learning phase (backpropagation) of the neural network using gradient descent optimization. In the backpropagation phase, the learnable parameters of the network are updated based on the gradient of the loss function. So, for any network, if $L_v = \frac{\partial L}{\partial v}$ is the incoming loss at the variance layer then the backpropagated loss from this layer with respect to the input $g(t)$, $L_{g(t)}$, is given by, 
\begin{align}
    L_{g(t)} &= \frac{\partial L}{\partial g(t)} = \frac{\partial L}{\partial v}  \cdot \frac{\partial v}{\partial g(t)}  = L_v \frac{2}{T}(g(t) - \mu)  
    \label{eqn:varDer} \\
\therefore \;\; L_{g(t)}  &\propto g(t) - \mu \nonumber
\end{align}

As it can be observed from (\ref{eqn:varDer}), the backpropagated loss from the Variance layer is proportional to the deviation of $g(t)$ from the mean of the signal. Therefore, the Variance layer provides more importance to the signal points which are away from the mean by assigning a higher proportion of the incoming gradient to these points. This also aligns with the characteristics of EEG wherein the deviation from the mean, in a form of ERD or ERS is a distinct signature of MI. 

Considering the suitability of the proposed Variance layer for EEG data, it was used for effective temporal features extraction in FBCNet. The output of SCB is passed to a Variance layer and it computes the temporal variance of the individual time-series in non-overlapping windows of size $w$, as given by (\ref{varForward}) in the forward pass.

\begin{equation}
    x_{V}(i,j, k) = \dfrac{1}{w} \sum_{t = w*k}^{(k+1)*w-1}(x_{SCB}(i,j,t) - \mu(i, j, k))^2 \label{varForward}
\end{equation}
where, $\mu(i, j, k)$ is the temporal mean of $x_{SCB}(i,j,t)$ within the $k^{th}$ window.

As it can be observed, the application of Variance layer across the entire time-duration reduces the number of features from $(m \times N_b \times T)$ to $(m \times N_b \times T/w)$ resulting in a high degree of feature reduction. In this work, the window length, $w$, was set to be 1s. Furthermore, the effect of different values of $w$ on model performance was evaluated in ablation analysis.  

\subsubsection{Classification}
Finally, the features extracted by the Variance Layer are passed through a log activation and are then provided to an FC layer with linear activation. The output of the FC is then passed to the softmax layer to get the output probabilities of each class. The FC layer weights are also regularized  by using a maximum norm constraint of 0.5; $||w||^2 < 0.5$ (Weight-normalization).

\subsection{Evaluation Datasets}

\begin{table*}[htbp]
\sisetup{detect-all=true,detect-inline-weight=math}
\centering
  \begin{threeparttable}
    \caption{Description of evaluation datasets}
    \begin{tabular}{lcS[table-format=3.0, table-column-width=4.5em]cS[table-format=3.0, table-column-width=6em]cccc}
        \toprule
        Dataset & Subject Type & {\# of Subjects} & \# of Sessions  & {\# Trials/Session} & \# of Classes & \# of Channels & \# Time-points & Analysis \\
        \midrule
        BCIC-IV-2A Data & Healthy Subjects & 9     & 2     & 288   & 4     & 22    & 1000  & CV, HO \\
        OpenMBI Data & Healthy Subjects & 54    & 2     & 200   & 2     & 20    & 1000  & CV, HO \\
        Stroke Data: A & Stroke Patients & 37    & 1     & 160  & 2     & 27    & 1000  & CV \\
        Stroke Data: B & Stroke Patients & 34    & 1     & 160  & 2     & 27    & 1000  & CV \\
        \bottomrule

    \bottomrule
    \end{tabular}%
    \label{tab:datasets}%
    \begin{tablenotes}
      \small
      \item CV: 10-fold Cross Validation on session 1 data, HO: Hold Out - training with session 1 data and test with session 2 data.
    \end{tablenotes}
  \end{threeparttable}
\end{table*}

In the BCI domain, extreme inter-subject variability is observed in the classification performance of different algorithms. Therefore, for robust comparison of the performance, we tested the FBCNet on the following four diverse EEG-MI datasets:
\begin{enumerate}
    \item \textit{BCIC-IV-2A Data}: A 4 class MI data from BCI Competition IV Dataset 2A \cite{Tangermann2012}.
    \item \textit{OpenMBI Data}: A 2 class MI data from Korea University EEG dataset \cite{Lee2019}. 
    \item \textit{Stroke Data: A}: A 2 class MI vs rest dataset \cite{Ang2015}.
    \item \textit{Stroke Data: B}: A 2 class MI vs rest dataset \cite{Ang2014}.
\end{enumerate}

Among these datasets, the BCIC-IV-2A Data \cite{Tangermann2012} and the OpenMBI Data \cite{Lee2019} contain the EEG data collected from healthy people. Moreover, these two datasets are publicly available and many state-of-the-art classification algorithms have used these datasets as a benchmark. Since application to post-stroke motor rehabilitation is also the focus of this paper, we include two big datasets, Stroke Data: A and Stroke Data: B, containing the MI data from chronic-stroke patients in the analysis. The Stroke Data: A is collected as a part of a clinical trial that investigated the combined effect of BCI-mediated rehabilitation and transcranial Direct Current Stimulation (tDCS) in chronic stroke patients \cite{Ang2015}. The Stroke Data: B is collected as a part of another clinical trial investigating the efficacy of BCI-mediated upper extremity motor rehabilitation in chronic stroke patients \cite{Ang2014}. Altogether, the proposed architecture was evaluated on MI-EEG data from 63 healthy people and 71 stroke patients. All the analyses in this work were performed on 0-4 seconds post-cue data.  

The most important dataset characteristics are summarized in Table \ref{tab:datasets} and more detailed data description is provided in supplementary material section S2.   

\subsection{Experiments}
\label{Experiments}
To evaluate the performance of FBCNet a Cross-Validation (CV) and a Hold Out (HO) analyses were conducted. 

The CV analysis was conducted in a 10 fold setting, with the 9 folds being used for training and 1 fold for testing. The folds were constructed by a sequential, class-balanced allocation of trials and this allocation was maintained constant for the entire analysis. The complete data from Stroke Data: A and Stroke Data: B datasets, and session 1 data from BCIC-IV-2A and OpenBMI datasets were used in CV analyses. We did not use the inter-session data in the CV analysis to avoid the confounding influence of inter-session variability which is a known problem in the BCI domain. 

To understand the effect of inter-session variability on the classification performance we conducted a subject-specific, inter-session, HO analysis for the BCIC-IV-2A Data and OpenBMI Data. In HO analysis, the complete data from session 1 for the given subject was used for the training purpose, and the resulting model was tested on the session 2 data. This analysis provides information about algorithms' capabilities in extracting highly generalizable discriminative features which remain valid during inter-session classification. 

In this manner, the performance of FBCNet was compared with one classical machine learning algorithm of FBCSP-SVM \cite{Ang2008}, and two state-of-the-art CNN architectures namely, Deep ConvNet \cite{Schirrmeister2017}, and EEGNet-8,2 \cite{Lawhern2018}. All the methods were used in the most optimal settings as recommended by the respective authors, and the detailed settings for each of these algorithms are provided in the supplementary material section S3. 

\subsection{Training Procedure}
\label{TrainingProcedure}
The same training procedure was followed for FBCNet, Deep ConvNet, and EEGNet-8,2. Architectures were trained using Adam optimizer at default settings (learning rate = $0.001$, betas = $0.9, 0.999$ ) \cite{Kingma2014}. The log-cross-entropy loss was used for gradient updates. As proposed in \cite{Schirrmeister2017}, a two-stage training strategy was used wherein the training data was further divided into a training set and a validation set. In the first stage, the model was trained using only the training set with the early stopping criteria whereby the validation set accuracy was monitored and training was stopped if there was no increase in the validation set accuracy for consecutive 200 epochs. After reaching the stopping criteria, network parameters with the best validation set accuracy were restored \cite{Schirrmeister2017}. Starting from this model, the training procedure was continued in the second stage wherein the model was trained with the complete training data (train + validation set). The second stage training was stopped when the validation set loss reduced below the stage 1 training set loss. Lastly, to avoid the case of infinite training in the situation of non-convergence, the maximum number of training epochs were limited to 1500 and 600 for training stage 1 and 2 respectively. In the CV analysis, the data from  1 among the 9 training folds was separated as a validation set. In the HO analysis, 20\% of the training data was set aside as a validation set. In both the CV and HO analysis the test fold/set was never used in any of the training steps. 
 
\subsection{Interpretability and Visualizations}
\label{Iterp}
The adaption of deep learning in the medical domain necessitates some form of explanation for models' decisions. Therefore, to understand the EEG features which the FBCNet learned to pay attention to, and to explore the difference between the data from healthy subjects and stroke patients, a method of DeepLift with the Rescale rule \cite{Shrikumar2017, captum2019github} was employed. The DeepLift algorithm calculates the relevance of every input feature on the resulting classification decision for each trial. In this analysis, from the training data, we used an average of all trials belonging to one class as a reference to the DeepLIFT algorithm and computed the single-trial relevance for every trial from the other class. Next, the subject-level relevance of input signals was calculated by averaging the normalized absolute per-trial relevance over all the trials. The extracted relevance was then used to infer the importance given by the trained model to a particular set of EEG channels and frequency bands. Finally, to summarize the properties of data from stroke patients and healthy subjects, a dataset-level relevance analysis was conducted. Here, the trained FBCNet models form the subjects with classification accuracy $>70$\% from the OpenBMI Data and Stroke Data: A and Stroke Data: B (combined together as Stroke Data) were analyzed. We present the relevance scores from subjects with classification accuracy $>70$\% because they represent robustly trained models which have encoded the information that generalizes on the test data. Moreover, a small percentage of subjects are known to be BCI-illiterate, that is, these subjects can not generate class discriminative MI-EEG patterns and this is another reason for the exclusion of subjects with accuracy $<$70\% for the interpretability analysis \cite{Lee2019}. 

\subsection{Statistical Analysis}

The statistical significance of differences in classification accuracy achieved by different algorithms was assessed using a Wilcoxon signed-rank test for BCIC-IV-2A Data (small sample size) and paired t-test for the remaining three datasets. Furthermore, to control the family-wise error rate, p-values were corrected with Bonferroni correction for multiple comparisons.  

\begin{table*}[htbp]
\sisetup{detect-all=true,detect-inline-weight=math}
\centering
  \begin{threeparttable}
    \caption{Average subject-specific classification accuracy}
    \begin{tabular}{L{9em}C{12em}S[table-format=2.2, table-column-width=8em]S[table-format=2.2, table-column-width=8em]S[table-format=2.2, table-column-width=8em]S[table-format=2.2, table-column-width=6em]}
    \toprule
    {Dataset} & {Test Configuration} & {FBCSP-SVM} & {Deep ConvNet} & {EEGNet-8,2} & {FBCNet} \\
    \midrule
    BCIC-IV-2A Data & 10-fold cross validation & 75.89 & 72.20 & 73.13 & \bfseries 79.03 \\
    OpenBMI Data & 10-fold cross validation & 64.61** & 68.33** & 70.89 & \bfseries 74.70 \\
    Stroke Data: A & 10-fold cross validation & 71.37** & 68.81** & 69.15** & \bfseries 79.16 \\
    Stroke Data: B & 10-fold cross validation & 74.14** & 71.11** & 73.47** & \bfseries 81.11 \\
    BCIC-IV-2A Data & Hold out test set & 68.06* & 72.22 & 73.15 & \bfseries 76.20 \\
    OpenBMI Data & Hold out test set & 60.36** & 60.77** & 63.63 & \bfseries 67.19 \\
    \bottomrule
    \end{tabular}%
    \label{tab:mainResults}%
    \begin{tablenotes}
      \footnotesize
      \item  The best performing method for each analysis is highlighted in boldface. The *, and ** represent that the classification performance of FBCNet is significantly better than the given baseline method with *: $\mathrm{p_{corrected} < 0.05}$ and, **:$\mathrm{p_{corrected} < 0.01}$.
    \end{tablenotes}
  \end{threeparttable}
\end{table*}

\section{Results}

\subsection{Classification accuracy}

\resHeader{FBCNet achieved significantly better classification accuracies compared to baseline methods}
Table \ref{tab:mainResults} presents complete classification results for all datasets using all the methods. From Table \ref{tab:mainResults}, it can be observed that FBCNet achieved the best classification performance across all datasets in both CV and HO analysis. The maximum accuracy improvement of 10\% was observed between FBCNet and FBCSP-SVM for CV analysis on the OpenBMI dataset. Furthermore, the FBCNet achieved a new SOTA of 76.20\% 4 class classification accuracy on the BCIC-IV-2A Data in the HO settings.  Also, in most datasets, the improvement in accuracy achieved by FBCNet over baseline methods was statistically significant (Table \ref{tab:mainResults}). More detailed classification results are presented in the supplementary material section S4.

\resHeader{FBCNet matched the performance of best performing baseline method for most subjects}
\begin{figure*}[tbh]
      \centering
      \includegraphics[width=7in]{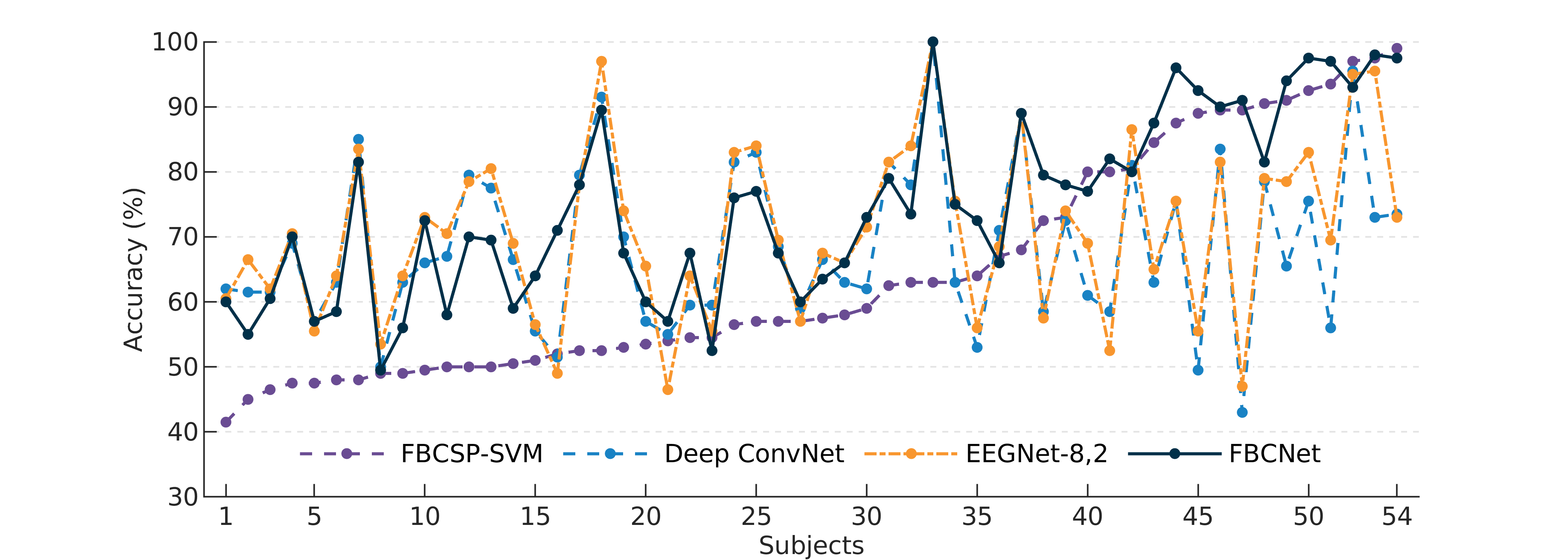}
      \caption{Classification accuracy for each subject from OpenBMI Data in 10-fold cross-validation settings (sorted by FBCSP-SVM acc.). It can be observed that deep learning architectures (Deep ConvNet, EEGNet-8,2) performed far better than FBCSP-SVM, the classical approach, for subjects with FBCSP-SVM accuracy $<$70\%. On the other end of the spectrum, the performance of deep learning architectures was far worse than that of FBCSP-SVM in subjects with FBCSP-SVM accuracy $>$70\%. Here, in contrast, FBCNet matched the performance of the best performing method for most of the subjects resulting in the best subject averaged classification accuracy. A similar trend was observed in other datasets as well.}
      \label{fig:classResults}
\end{figure*}
\begin{table*}[htbp]
  \centering
  \caption{Average classification accuracies for 25\% subjects with highest and lowest cross validation accuracy in each dataset}
    \begin{tabular}{lcccccccc}
    \toprule
    \multicolumn{1}{c}{\multirow{2}[4]{*}{Dataset}} & \multicolumn{4}{c}{Top 25\%}  & \multicolumn{4}{c}{Bottom 25\%} \\
\cmidrule{2-5} \cmidrule{6-9}          & FBCSP-SVM & Deep ConvNet & EEGNet-8,2 & FBCNet & FBCSP-SVM & Deep ConvNet & EEGNet-8,2 & FBCNet \\
    \midrule
OpenBMI Data & 90.11 & 84.79 & 87.43 & \textbf{93.75} & 48.00 & 54.50 & 54.61 & \textbf{57.60} \\
    Stroke Data: A & 90.46 & 83.71 & 84.96 & \textbf{95.06} & 53.23 & 53.74 & 52.28 & \textbf{59.52} \\
    Stroke Data: B & 89.53 & 82.05 & 84.21 & \textbf{93.14} & 53.55 & 61.88 & 65.00 & \textbf{67.22} \\
    \bottomrule
    \end{tabular}%
  \label{tab:topBot25}%
\end{table*}%

Following the analysis of classification accuracies averaged over all subjects, the performance of all algorithms for every subject was investigated. As an example, Fig. \ref{fig:classResults} presents the 10-fold CV accuracies for all subjects in OpenBMI Data. Here, a significant difference was observed in the classification accuracies achieved by different algorithms for each subject. In particular, both baseline deep learning methods, Deep ConvNet, and EEGNet-8,2, resulted in similar accuracies, which were significantly different from the accuracies achieved by the classical machine learning approach of FBCSP-SVM (paired t-test, $\mathrm{p <0.05}$). Moreover, it was observed that deep learning architectures performed far better than the classical approach, for subjects with FBCSP-SVM accuracy $<$70\%. Conversely, the performance of deep learning architectures was worse than that of FBCSP-SVM in subjects with FBCSP-SVM accuracy $>$70\%. In contrast to all baselines, as can be observed from Fig. \ref{fig:classResults}, FBCNet matched the performance of the best performing method for most subjects resulting in the best average classification accuracy. A similar but statistically insignificant trend was observed in other datasets as well.

To quantitatively investigate this pattern, we analyzed the average classification accuracy for 25\% subjects with the highest and lowest CV accuracy in each dataset (Table \ref{tab:topBot25}). Here, the average classification accuracies of top 25\% subjects for FBCSP-SVM were observed to be significantly higher than that of Deep ConvNet, and EEGNet-8,2 (t-test, all p $<$ 0.05 except $\mathrm{p_{FBCSP-EEGNet}}$ for OpenBMI Data) and the accuracy for FBCNet was even better than that of FBCSP-SVM. For the bottom 25\% subjects, FBCSP-SVM was observed to perform worse than deep learning algorithms  (all p $<$ 0.01 except Stroke Data A) and the FBCNet achieved the highest average accuracy for this class of subjects as well. 

\resHeader{Highest number of subjects achieved $>$70\% accuracy with FBCNet}
In the BCI field, a system with $>$ 70\% 2-class classification accuracy is generally considered to be usable by healthy subjects and stroke patients \cite{shu2018fast}. Therefore, the number of subjects for whom the classification algorithm managed to achieve at least 70\% CV accuracy were analyzed and the results are presented in Table \ref{tab:nAcc70}. Among baseline methods, EEGNet-8,2 resulted in the most number of subjects being able to achieve $>$70\% accuracy for OpenBMI Data. Contrarily, for both stroke datasets, FBCSP-SVM resulted in the most number of subjects with $>$70\% accuracy. Better than all baseline methods, for all datasets, FBCNet was able to achieve $>$70\% accuracy for the most number of subjects. Compared to the best baseline method (FBCSP-SVM), FBCNet resulted in 28\% more stroke patients (Stroke Data: A+B) with $>$70\% accuracy. 

\begin{table}[tbp]
  \centering
  \caption{Number of subjects with CV classification accuracy $>$ 70\%}
    \begin{tabular}{lC{3em}C{3em}C{3em}C{3em}C{3em}}
    \toprule
    Dataset & Total Subjects & FBCSP-SVM & Deep ConvNet & EEGNet-8,2 & FBCNet \\
    \midrule
OpenBMI Data & 54    & 17    & 21    & 27    & \textbf{33} \\
    Stroke Data: A & 37    & 20    & 18    & 19    & \textbf{28} \\
    Stroke Data: B & 34    & 23    & 16    & 22    & \textbf{27} \\
    \midrule
    Stroke Data (A+B) & 71    & 43    & 34    & 41    & \textbf{55} \\
    \bottomrule
    \end{tabular}%
  \label{tab:nAcc70}%
\end{table}%

\resHeader{FBCNet was least affected by datasets with fewer training samples}
\begin{figure}[tb]
      \centering
      \includegraphics[width=3.5in]{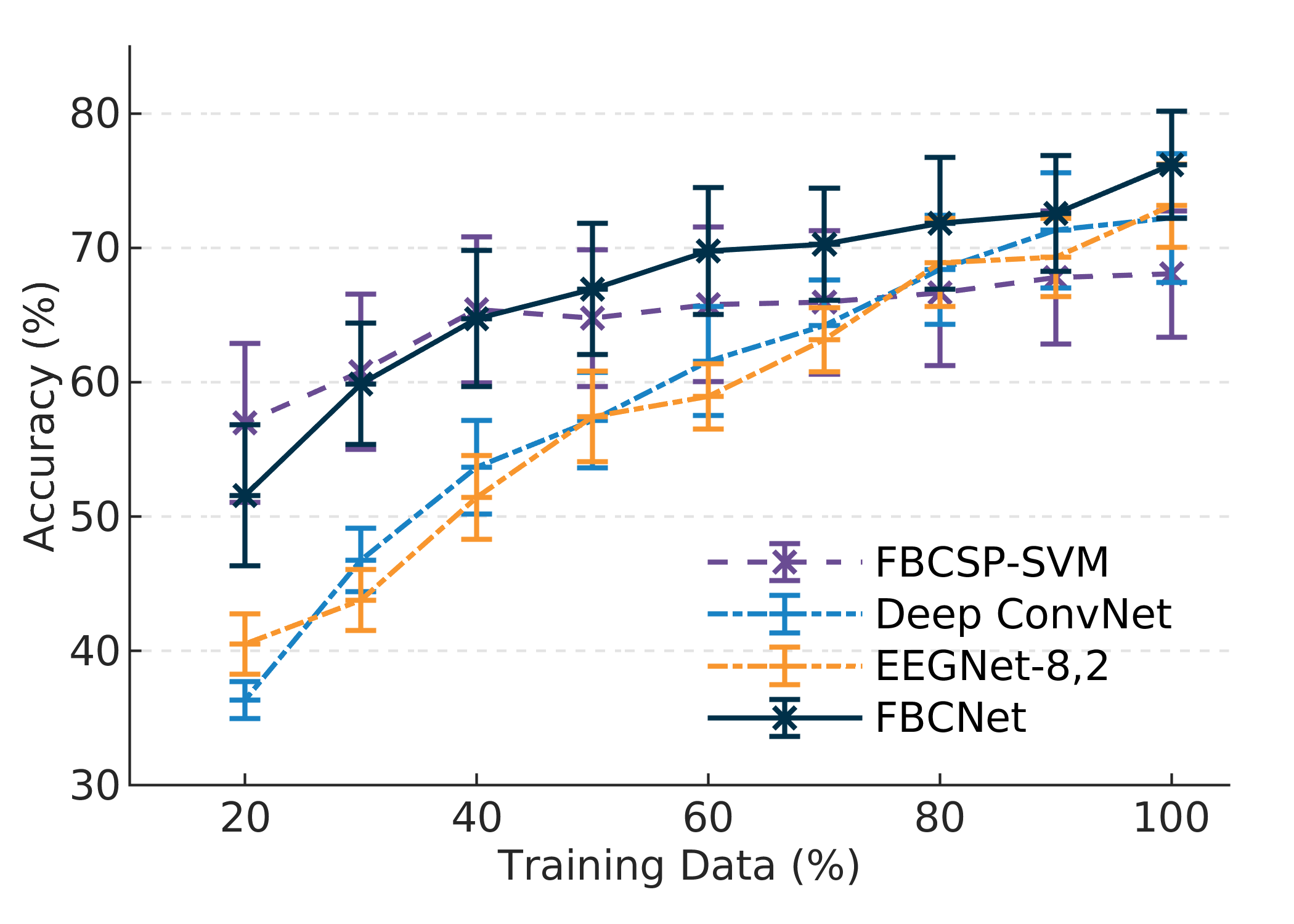}
      \caption{The sensitivity of classification algorithms to small training sets. Here, the effect of a small amount of data on the test accuracy is evaluated using the BCIC-IV-2A Data for various algorithms. A fraction of the training data (x-axis) is used to train a model, which is tested on data from an independent test session. It can be observed that the baseline deep learning architectures (Deep ConvNet and EEGNet-8,2) are highly sensitive to the small training sets, whereas the classical approach of FBCSP-SVM is relatively less susceptible. The proposed method (FBCNet) matches the accuracy achieved by deep learning methods in the presence of ample training data while retaining relatively better performance even when the training set is small. The error bars represent a standard mean error.}
      \label{fig:smallSet}
\end{figure}

The sensitivity of the classification algorithm to fewer training samples was evaluated using the separate test session data from BCIC-IV-2A Data in the HO analysis. Here, all the algorithms were trained using various fractions of the training dataset ranging from 20\% to 100\% in steps of 10\%. Trained models were then tested on a separate session 2 data and the test accuracy was analyzed as a function of training data percentage. Results of this analysis are presented in Fig. \ref{fig:smallSet}. Here, both baseline deep learning architectures were observed to be highly sensitive to small training sets with an accuracy drop of almost 35\% when the number of training trials was reduced to 20\%. Contrarily, FBCSP-SVM was least affected by a reduction in the training set. In fact, the drop in accuracy using FBCSP-SVM only became statistically significant when the portion of training samples was reduced to 30\%. Furthermore, compared to both Deep ConvNet and EEGNet-8,2, better classification accuracies achieved by FBCSP-SVM were statistically significant when training data was reduced below 50\%. However, the maximum accuracy achieved by FBCSP-SVM, when using the complete training data was much lower compared to these baseline deep learning methods (Table \ref{tab:mainResults}). Different from FBCSP-SVM, at 100\% of the training data, the best accuracy was achieved by FBCNet. Moreover, the accuracy curve of FBCNet closely followed the characteristics of FBCSP-SVM with a similar sensitivity to the reduced training set as that of FBCSP-SVM.

\resHeader{Temporal feature extraction using Variance layer resulted in significantly better classification accuracies}
\begin{figure}[tb]
      \centering
      \includegraphics[width=3.5in]{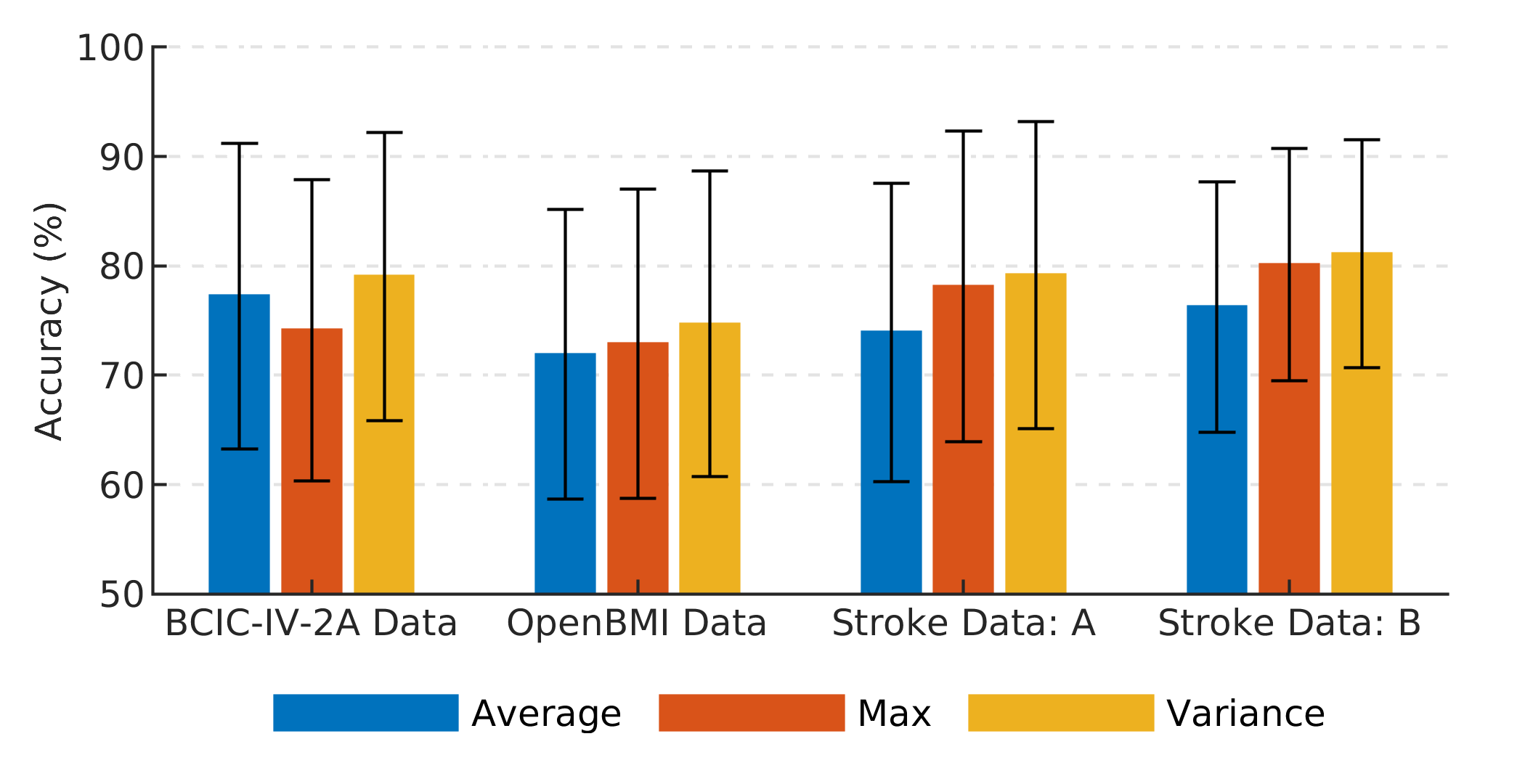}
      \caption{FBCNet cross validation classification accuracies with different temporal feature extraction layers (mean$\pm$std). Temporal feature extraction using Variance layer resulted in best classification accuracies across all datasets.}
      \label{fig:bar-temporalLayer}
\end{figure}

To analyze the contribution of the novel Variance layer in the improved results achieved by FBCNet, we investigated the effect of different temporal feature extraction layers on classification accuracies and Fig. \ref{fig:bar-temporalLayer} presents these results. Among all the temporal feature extraction layers, FBCNet with the Variance layer achieved the highest classification accuracies in all analyses. Feature extraction using the Average and Max layer resulted in consistently worse accuracies across all datasets, and compared to the Variance layer, these differences were statistically significant (all p $<$ 0.05 expect for BCIC-IV-2A Data).

\resHeader{Increasing the number of spatial filters and reducing the window size in FBCNet architecture resulted in marginally improved classification results}
\begin{figure}[tb]
      \centering
      \includegraphics[width=3.5in]{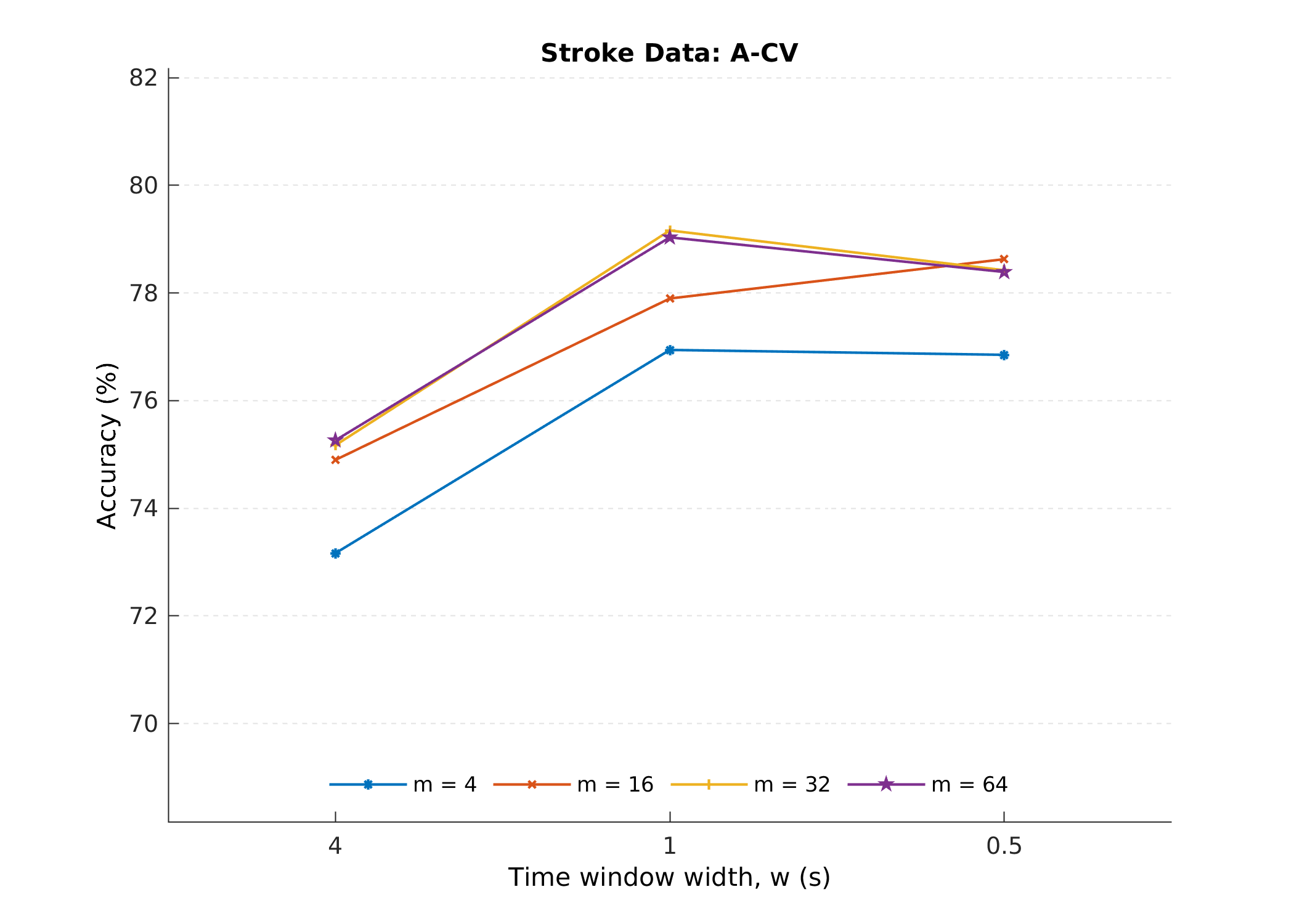}
      \caption{FBCNet cross-validation classification accuracies with different number of spatial filters per frequency band ($m$) and variance window length ($w$).}
      \label{fig:bar-m}
\end{figure}

The average classification accuracy achieved by FBCNet on the Stroke Data: A with a different number of spatial filters and window lengths is presented in Fig. \ref{fig:bar-m}. In this analysis, increasing the number of spatial filters and decreasing the window size initially produced increased accuracy. However, too many spatial filters and too small window sizes resulted in reduced classification accuracy. Also, the computational complexity of FBCNet was linearly proportional to the number of spatial filters and inversely proportional to the window size. Therefore, changes in these parameters resulted in FBCNet with more learnable parameters and longer training times.

\begin{table}[tbp]
  \centering
  \caption{Comparison of FBCNet with existing methods for the BCIC-IV-2A Data (4-class classification \%Acc. (kappa value)). }
    \begin{tabular}{lr@{\hskip 2pt}lr@{\hskip 2pt}l}
    \toprule
    Algorithm & \multicolumn{2}{c}{CV} & \multicolumn{2}{c}{Test session} \\
    \midrule
    FBCSP \cite{Ang2012c} & 74.72 &(0.663) & 67.75&(0.569) \\
    C2CM \cite{Sakhavi2018} & 78.78 &       & 74.46&(0.659) \\
    Deep ConvNet \cite{Schirrmeister2017} & -     &       & 70.90 &  \\
    EEGNet \cite{Lawhern2018} & 70.00 &       & -     &  \\
    RFNet \cite{Zhang2020} & -     &       & 75.51 & (0.673) \\
    MI-EEGNet \cite{Riyad2021} & 77.49 &       & 74.61 &  \\
    \midrule
    \textbf{FBCNet (proposed)} & \textbf{79.03} & \textbf{(0.720)} & \textbf{76.20} & \textbf{(0.683)} \\
    \bottomrule
    \end{tabular}%
  \label{tab:addlabel}%
\end{table}%

\subsection{Interpretability and visualizations}

\resHeader{Greater inter-subject variability in the relevance patterns was observed for stroke patients}

\begin{figure*}[tbh]
      \centering
      \includegraphics[width=7in]{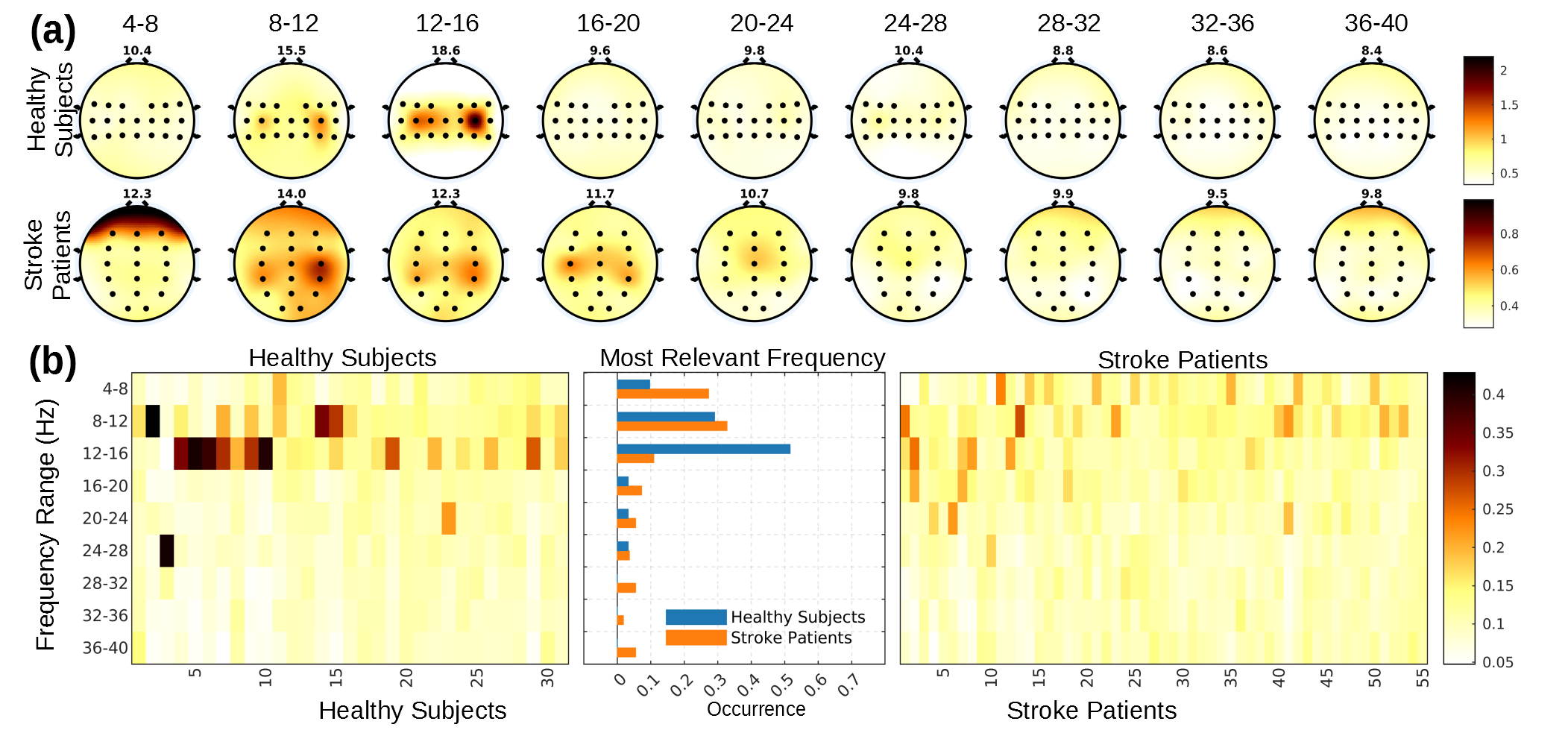}
      \caption{Group level relevance analysis for healthy subjects and stroke patients. Part (a) presents the subject-averaged channel-frequency input relevance patterns for healthy subjects and stroke patients. The number in the title is the percentage relevance for the given frequency band. The input relevance is highly concentrated in the 8-16Hz frequency range at the C3 and C4 channels for healthy subjects whereas the relevance patterns for stroke subjects are highly diffused. Investigating more on the subject-averaged relevance, Part (b) presents the heatmap of frequency band relevance for each subject. Also, the histogram in the center presents a normalized count of subjects for whom the given frequency band has the highest relevance. For healthy subjects, the 8-16 Hz frequency range has consistently high relevance whereas the most relevant frequency range largely differed across stroke patients.}
      \label{fig:groupInterpTop25}
\end{figure*}

To understand general trends in the relevance patterns learned by the FBCNet model across subjects and to explore if the data from stroke patients is any different from healthy subjects, a group-level relevance analysis we performed and its results are presented in Fig. \ref{fig:groupInterpTop25}. 

First, subject-averaged relevance patterns (Fig. \ref{fig:groupInterpTop25} (a)) at all frequency bands and EEG channels for stroke patients' and healthy subjects' data were inspected. For healthy subjects, the 12-16Hz and 8-12Hz were observed to be the two most relevant frequency bands and they constituted 34\% of the total input relevance averaged across all subjects. Also, the channel relevance in these two frequency bands was most concentrated at the left and right motor areas of the brain (C3, C4). All these characteristics were closely associated with the known MI signatures. In stroke patients, the averaged relevance patterns were observed to be more diffused and all the frequency bands in the 4-24Hz range resulted in similar input relevance. Moreover, the channel relevance patterns in these frequency bands were also much more diffused and many channels received similar total relevance scores. Yet, the C4, CP4 and, P4 channels in the 8-12Hz range, C3, C4, and CP4 channels in the 12-16Hz range, and F7 and F8 channels in the 4-8Hz range, were observed to have slightly higher relevance than other channels. 

Next, the channel-frequency band relevance patterns for each healthy and stroke subject were visually inspected. Here, for most healthy subjects, the 8-12 Hz frequency band was observed to be highly relevant whereas the most relevant frequency range largely differed across stroke patients. To concisely visualize this difference, a heat-map of filter band relevance for healthy and stroke subjects was plotted and it is presented in \ref{fig:groupInterpTop25} (b). From the heatmap, as well as the normalized histogram, the 12-16Hz was observed to be the frequency band with the highest relevance in half of the healthy subjects. Contrarily, for stroke patients, no single highly relevant frequency band could be identified, and the most relevant frequency band was highly subject-specific. Moreover, for each stroke patient, the input relevance was distributed across multiple frequency bands, and the difference in the relevance of the first and the second most relevant frequency band was quite low. Furthermore, this difference was significantly different from healthy subjects' data (independent samples t-test, p $<$ 0.05). 

\section{Discussion}
In this paper, we proposed FBCNet, a novel, neurophysiologically inspired, end-to-end CNN architecture for MI classification that can learn generalizable discriminative features in the presence of limited data and produce better classification accuracies. We evaluated FBCNet against the state-of-the-art classical machine learning and deep learning approaches using EEG-MI data from a large corpus of healthy subjects (n = 54 $+$ 9 = 63), and chronic stroke patients (n = 37 $+$ 34 = 71).  To the best of our knowledge,  this is the first work that compares the performance of deep learning approaches with classical machine learning methods for such a large population of healthy and stroke subjects. Besides the differences in the classification accuracies, using interpretability analysis, this work also presented some of the key differences in stroke patients and healthy subjects data from the perspective of MI classification. Next, we showed that the use of a hybrid approach, as done in FBCNet, which leverages the complex feature learning capabilities of deep learning methods and mitigates their sensitivity to small datasets by incorporating the neurophysiological priors for MI classification in the architecture design may lead to significant improvements in MI-BCI classification accuracies. We contemplate that, the three-stage approach of the spectral, spatial, and temporal localization of EEG features in FBCNet, has resulted in an architecture that is constrained enough that it effectively focuses on neurophysiological signatures of MI even in the absence of large training data while being flexible enough to efficiently handle the intra-class variability in EEG trials. Therefore, striking a balance between model complexity (more trainable parameters to extract deeper encoded generalizable patterns) and constraints (neurophysiologically reasonable patterns) can be the way to successfully adapt deep learning methods into the BCI domain. 

\subsection{Performance of classical machine learning methods and state-of-the-art deep learning architectures}
In our experiments, there was no statistically significant difference between the subject-averaged classification accuracy achieved by FBCSP (classical machine learning method) and Deep ConvNet, EEGNet-8,2 (deep learning methods) for all but one analysis and these results are in line with the previous literature \cite{Schirrmeister2017, Lawhern2018, Raza2020}, indicating that the deep learning architectures are yet to present substantial improvements in the EEG domain. Furthermore, although statistically insignificant, FBCSP was observed to be the best performing baseline method for both stroke subjects' datasets and a reversed trend was observed for the healthy subjects' data from the OpenBMI dataset. These results may have been caused by different amounts of training data present across these datasets. Both stroke datasets had 20\% fewer training trials as compared to the OpenBMI dataset (160 vs 200 trials/subject) and this reduction in training data may have affected the performance of deep learning architectures on a larger scale as compared to its impact on the FBCSP. This hypothesis was also further supported by the results of variable training data analysis performed on the BCIC-IV-2A Data (Fig. \ref{fig:smallSet}) wherein, compared to FBCSP, the accuracy of existing deep learning methods dropped significantly when trained with the reduced amount of data. This indicates that in the absence of adequate training data, the general-purpose deep learning architectures may result in worse classification accuracies. 

Furthermore, although, the difference in the amount of training data can account for dataset-level differences in classification accuracies of the baseline architectures, it can not explain the large differences in classification performance of these algorithms for individual subjects. As observed from Fig. \ref{fig:classResults}, for most subjects, the classification accuracy differed significantly between FBCSP and baseline deep learning methods. Here, we observed a general pattern that FBCSP resulted in extreme (very high or low) classification accuracies for most subjects, and contrarily, average classification accuracies were achieved by most subjects using Deep ConvNet and EEGNet-8,2. Inter-subject variability in the EEG data distributions can be one possible explanation for these observed differences. It is probable that the neurophysiologically constrained model of FBCSP is inadequate for subjects with high intra-class variability and non-standard MI-EEG patterns, and more complex deep learning models may achieve relatively better accuracy for these subjects by learning some of those complex patterns. However, this additional flexibility in the deep learning architectures may also be detrimental to their performance in subjects with better class discriminative features. In this case, particularly with the limited and noisy training data, deep learning architectures may end-up learning non-generalizable patterns leading to poor test set performance. This indicates that striking a balance between model complexity and constraints may benefit all BCI users.
 
\subsection{Classification performance of FBCNet}

FBCNet was designed to strike the balance between model capacity and complexity by incorporation of neurophysiological knowledge within the deep learning framework and this principle was hypothesized to results in a model that has optimal learning capacity while being less susceptible to relatively small and noisy EEG data. The results using FBCNet strongly support this hypothesis wherein the FBCNet has resulted in best classification accuracy across all datasets and this improvement in accuracy by FBCNet over baseline methods is statistically significant for most datasets (Table \ref{tab:mainResults}). Furthermore, in the analysis of individual subject classification accuracies, we observed that FBCNet closely matched the performance of the best performing classification algorithm for most subjects in the datasets (Fig. \ref{fig:classResults}). Due to this, the average classification accuracy of 25\% best performing subjects by FBCNet was even better than that of FBCSP (Table \ref{tab:topBot25}). Furthermore, the FBCNet achieved the best average classification accuracy for worst performing 25\% subjects (Table. \ref{tab:topBot25}), and this consequently resulted in the most number of subjects being able to control the BCI with $>$70\% classification accuracy using FBCNet (Table \ref{tab:nAcc70}). Furthermore, we observed that the FBCNet is much less susceptible than baseline deep learning architectures to small training sets (Fig. \ref{fig:smallSet}) indicating that the neurophysiologically constrained architecture of FBCNet can more effectively learn the generalizable class-discriminative patterns and avoid overfitting in cases of limited training data. Moreover, unlike FBCSP, in the presence of ample training data, the FBCNet accuracies improved continuously indicating that FBCNet can also effectively utilize a higher amount of training data when it is available. All these results indicate that the incorporation of neurophysiological knowledge within the deep learning framework, as done in FBCNet, may lead to better MI classification accuracies. Furthermore, a similar observation can be made from the existing literature wherein RFNet \cite{Zhang2020}, which fuses Riemannian geometry with deep learning methods, held the record for best accuracy on the BCIC-IV-2A to date. 

\subsection{Role of Variance Layer}

The Variance layer for temporal feature extraction is another important contribution of this paper. The use of variance operation along the temporal dimension was motivated by the fact that variance of a filtered signal represents the spectral power in the time-series, and spatio-temporal differences in EEG spectral power are known class-discriminative features of MI. Therefore, we hypothesized that variance may be a more suitable operation for temporal consolidation of EEG signals and this hypothesis was confirmed by experimental results. In the ablation analysis of temporal feature extraction layer, we observed that FBCNet with Variance layer resulted in significantly better performance than Average and Max layer which are the two most widely used feature reduction strategies in the deep learning field (Fig. \ref{fig:bar-temporalLayer}). Hence Variance layer might be better suited for temporal feature consolidation in deep learning networks designed for MI classification.

\subsection{Effect of hyperparameter selection}

The number of spatial filters per frequency band ($m$) and length of the variance window ($w$) are the two most important hyperparameters in the FBCNet architecture. These two parameters respectively control the spatial and temporal feature representation capacity of FBCNet, and together they set the computational complexity and model capacity of FBCNet. Therefore, it is important to properly set the values of these parameters. Here, we observed that the $m$ between 16 and 32 and $w$ between 0.5s and 1s led to optimal performance on most datasets. Also, it was observed that more complex tasks like 4 class classification in BCIC-IV-2A Data, benefit from larger $m$ values, indicating that such tasks need a higher model capacity to correctly learn the more complex data and FBCNet can easily accommodate this requirement. The observed optimal value of $w$ between 0.5s and 1s is also neurophysiologically sound wherein the ERD/ERS patterns associated with MI are known to have a duration of 0.5s to 1s.

\subsection{Interpretability Analysis}
Interpretability of the knowledge learned by the neural network architecture is a necessity when used in clinical settings. In this work, we observed the FBCNet models learned to assign relevance to the EEG channels in the motor region of the brain (C3 and C4) in the 8-16Hz and 24-32Hz frequency range for healthy subjects (Fig. \ref{fig:groupInterpTop25}). This indicates that FBCNet could successfully focus on the neurophysiologically sound features because spectro-spatially discriminative patterns at the primary motor cortex in the alpha (8-12Hz) and high beta bands (23-30Hz) of EEG are well-known signatures of left vs right-hand MI \cite{Pfurtscheller2006}. Moreover, we observed that by encapsulating the true signatures of MI, these features also generalized well on the unseen test data which explained the higher classification accuracies achieved by FBCNet. 

In the interpretability analysis, we also investigated the possible differences in the class discriminative patterns between healthy subjects and stroke patients. Here, we observed consistent relevance patterns across multiple subjects with the highest relevance in the 8-16Hz and 24-32Hz frequency range at the motor region of the brain (C3, C4 channels) for healthy subjects data. These relevance patterns are in-line with the neurophysiological signatures of MI \cite{Pfurtscheller2006}. However, for stroke patients, these patterns showed large inter-subject variability, and they were also much more spatially and spectrally diffused for individual patients. These spectro-spatially spread relevance patterns in stroke patients may indicate that the damage in the brain caused by stroke may have resulted in compensatory recruitment of non-motor areas by the brain \cite{Cramer2008a}. However, to further investigate this phenomenon, more data with distinct MI tasks from stroke patients is needed. Yet, the large inter-subject differences between the stroke patients' relevance pattern can be attributed to stroke-induced subject-specific modification in the brain, and therefore, the compensatory recruitment of non-motor areas is a plausible explanation for the spectro-spatially spread relevance patterns in stroke patients. 

\section{Conclusion}
This paper proposed a neurophysiologically motivated end-to-end CNN architecture named FBCNet for subject-specific MI classification that can learn generalizable discriminative features in the presence of limited data and produce better classification accuracies. In this architecture, we proposed a hybrid approach for the task of MI classification in which we leveraged the complex feature learning capabilities of deep learning methods and mitigated their sensitivity to small datasets by incorporating the neurophysiological priors of MI in the architecture design. This proposed approach achieved significantly better classification accuracies across four motor imagery datasets among which two were collected from chronic stroke patients. Moreover, with interpretability analysis, we demonstrated that improved performance by FBCNet was driven by the effective learning of neurophysiologically relevant EEG features. Furthermore, we showed that there are differences in the MI data between healthy subjects and stroke patients, and FBCNet can perform well on both healthy and patient data. Overall, the results indicate that inclusion of neurophysiological priors while designing deep learning architectures, as done in this work, will result in an architecture that is constrained enough that it can effectively focus on neurophysiological signatures even in the absence of large training datasets while being complex enough to effectively utilize a higher amount of training data when it is available, and such an approach may lead to better classification results in the field of MI-BCI.  

\section*{Acknowledgment}
This work was partially supported by the RIE2020 AME Programmatic Fund, Singapore (No. A20G8b0102), and the  Institute of Information \& Communications Technology Planning \& Evaluation (IITP) grant funded by the Korea government (No. 2017-0-00451, and No. 2019-0-00079). The computational work for this article was partially performed on resources of the National Supercomputing Centre, Singapore.

\bibliographystyle{IEEEtran}

\bibliography{ref.bib}
\end{document}


\setcounter{equation}{0}
\setcounter{figure}{0}
\setcounter{table}{0}
\setcounter{page}{1}
\makeatletter
\renewcommand{\theequation}{S\arabic{equation}}
\renewcommand{\thefigure}{S\arabic{figure}}
\renewcommand{\thetable}{S\arabic{table}}
\renewcommand{\thesection}{S\arabic{section}}

\begin{titlingpage}
\maketitle
\end{titlingpage}

\section{FBCNet architecture}

The design philosophy of FBCNet is explained in the main manuscript section II. Here, we provide implementation level details of FBCNet in Table \ref{tab:architecture}.

\begin{table*}[htbp]
\sisetup{detect-all=true,detect-inline-weight=math}
\centering
  \begin{threeparttable}
    \caption{FBCNet architecture}
    \scriptsize
  \begin{tabular}{lp{12em}lllll}
    \toprule
    \multicolumn{1}{p{8.5em}}{Block} & Layer & \multicolumn{1}{l}{\# Kernels} & \multicolumn{1}{l}{Kernel Size} & \multicolumn{1}{l}{\# Parameters} & Output & \multicolumn{1}{p{15.5em}}{Options} \\
    \midrule
    \multicolumn{1}{p{8.5em}}{Spectral Filtering } & Input &       &       &       & (1, $C$, $T$) &  \\
          & Filter-Bank &       &       &       & ($N_b$, $C$, $T$) &  \\[0.2cm]
    \multicolumn{1}{L{8.5em}}{Spatial Convolution } & Depthwise Conv2D & \multicolumn{1}{l}{$m*N_b$} & \multicolumn{1}{l}{($C$, 1)} & \multicolumn{1}{l}{$m*N_b$*$C$ $+$ $m*N_b$} & ($m*N_b$, 1, $T$) & \multicolumn{1}{p{15.5em}}{depth = $m$, max L2 weightnorm = 2} \\
          & BatchNorm &       &       & \multicolumn{1}{l}{2*$m*N_b$} & ($m*N_b$, 1, $T$) &  \\
          & Activation - ELU &       &       &       & ($m*N_b$, 1, T) & \multicolumn{1}{p{15.5em}}{ELU, ReLU, Leaky-ReLU} \\[0.2cm]
    \multicolumn{1}{L{8.5em}}{Temporal Feature Extraction} & \multicolumn{1}{l}{Variance Layer} &       & \multicolumn{1}{l}{(1, $w$)} &       & ($m*N_b$, 1, $T/w$) & \multicolumn{1}{p{15.5em}}{Variance, Average, Max} \\[0.45cm]
    \multicolumn{1}{p{8.5em}}{Classifier}      & Flatten  &       &       &       & $m*N_b*T/w$ &  \\
     & Fully connected Layer & \multicolumn{1}{l}{$N_c$} &       & \multicolumn{1}{l}{$m*N_b*T/w*N_c+N_c$} & $N_c$ & \multicolumn{1}{p{15.5em}}{max L2 weightnorm = 0.5} \\
          & Log-Softmax &       &       &       & $N_c$ &  \\
    \bottomrule
    \end{tabular}%

    \label{tab:architecture}%
    \begin{tablenotes}
      \scriptsize
      \item $C$: number of EEG channels, $T$: number of time points, $N_b$: number of frequency bands, $m$: number of convolution filters per frequency band, $N_c$: number of output classes, $w$: temporal window length
    \end{tablenotes}
  \end{threeparttable}
\end{table*}

\section{Details of Evaluation Datasets}
The performance of FBCNet was tested on following four diverse datasets:
\begin{enumerate}
    \item \textit{BCIC-IV-2A Data}: A 4 class MI data from BCI Competition IV Dataset 2A \cite{Tangermann2012}.
    \item \textit{OpenBMI Data}: A 2 class MI data from OpenBMI dataset \cite{Lee2019}. 
    \item \textit{Stroke Data: A}: A 2 class MI vs rest dataset \cite{Ang2015}.
    \item \textit{Stroke Data: B}: A 2 class MI vs rest dataset \cite{Ang2014}.
\end{enumerate}
Further details of these datasets are as follows:

\label{S:DataDetils}
\subsection{BCIC-IV-2A Data}
The BCIC-IV-2A Data consists of EEG data from 9 subjects collected over 2 sessions. The data contains 4 classes; MI of left, and right hands, feet, and tongue. The EEG data has been recorded using 22 electrodes with a sampling frequency of 250 Hz. There are 72 trials of every class in each session and every trial is of 4s duration. In our analysis, we used all the 22 channels the entire 4 seconds data.

\subsection{OpenBMI Data}
OpenBMI Data contains 2 sessions of 2-class EEG-MI data from 54 healthy subjects \cite{Lee2019}. The MI of left and right hand are the two classes in the data and there are in total 100 trials of each class per session and each trial is 4s in length. The EEG data has been originally recorded at 1000Hz using 62 electrodes. In this analysis, as it is done in the original work \cite{Lee2019}, we have selected 20 channels in the motor region for the classification task (FC-5/3/1/2/4/6, C-5/3/1/z/2/4/6, and CP-5/3/1/z/2/4/6). Moreover, one of the baseline architectures, the Deep ConvNet is designed for the EEG data sampled at 250 Hz. Therefore, to maintain the compatibility and fairness of comparison, we down-sampled the data by a factor of 4 to have a sampling frequency of 250 Hz. Moreover, this analysis setting results in approximately similar EEG trial dimensions ($ channels  \; \times \; time$) across all the datasets.    

\subsection{Stroke Data: A}
Stroke Data: A is part of post-stroke BCI motor rehabilitation trial wherein 37 stroke patients are screened for their ability to control MI-BCI \cite{Ang2015}. The MI of stroke-paralyzed hand and rest are the two classes in the data and there are in total 80 trials of each class for every patient. Every trial is of 4 seconds duration and the EEG data is collected using 27 channels sampled at 250Hz with a hardware bandpass filtering from 0.5 -40Hz. In the present analysis, we use the entire 27 channel, 4s data for classification. 

\subsection{Stroke Data: B}
Stroke Data: B is also a part of post-stroke BCI motor rehabilitation trial and similar to Stroke Data: A, contains a 2-class data with MI of stroke-paralyzed hand being one class and rest being the other class \cite{Ang2014}. This dataset comprises of data from 34 stroke patients with 160 trials per patient. The data collection protocol is the same as that of the Stroke Data: A with the 4s of EEG data collected per trial using 27 channels sampled at 250 Hz. For this dataset as well, we use the entire 27 channel, 4s data for classification.  

\section{Implementation of Baseline Classification methods}
\label{S:BaseArch}

\subsection{ The Traditional Approach: FBCSP-SVM} 
We implemented the FBCSP-SVM algorithm as per its authors' recommendations in \cite{Ang2008}. As done in the original work \cite{Ang2008}, we decomposed the raw EEG data using 9 narrow-band bandpass Chebyshev Type II filters, each of 4Hz bandwidth, spanning from 4 to 40 Hz (4-8, 8-12, ..., 36-40 Hz) with transition bandwidth of 2Hz and stopband ripple of 30dB. The narrow-band EEG was then spatially filtered using CSP algorithm and 4 most discriminative CSP filters from each band where extracted. The log-variance of the EEG data filtered using CSP filters was extracted as a feature. From these 36 features (9 frequency bands $\times $ 4 features per band) best 8 features were selected using the MIBIFPW algorithm \cite{Ang2012d}. An SVM classifier was then trained using the selected features to classify the trial into one of the two classes. An epsilon - support vector regression with a radial basis function kernel flavour of SVM was used for the classification. 

The CSP algorithm is designed for a binary classification problem. Therefore, to classify the 4 class BCIC-IV-2A Data we used a One-Verses-Rest (OVR) strategy wherein 4 binary classifiers were trained to classify one class from the remaining 3 classes. Then each trial was assigned to the class for which it received the maximum SVR score among all 4 classifiers. Due to this approach, for BCIC-IV-2A Data, in total 16 CSP filters were extracted for each band (4 OVR-models $\times$ 4 filters per model).

Lastly, since both the FBCSP and FBCNet use the multi-view filtered EEG representation, we used the same filter bank in both of these algorithms for fairness in comparison. 
    
\subsection{ Existing CNN architectures : Deep ConvNet and EEGNet}
We implemented the Deep ConvNet \cite{Schirrmeister2017} and EEGNet-8,2 \cite{Lawhern2018} following the descriptions found in the respective papers. One major modification was done to the EEGNet-8,2 architecture due to the difference in the sampling frequencies of the data. The original EEGNet was proposed for the data with 128 Hz sampling frequency. However, all 4 datasets in this work have a sampling frequency of 250Hz. Therefore we multiply the lengths of temporal kernels and temporal pooling layers in their architectures by 2 to correspond approximately to the sampling rate in our model. Deep ConvNet and EEGNet-8,2 were trained in the exact same manner as that of the FBCNet. The exact implementations of these architectures used in this work are available at \url{https://github.com/ravikiran-mane/FBCNet}.

\section{Single subject classification accuracies}
\label{S:mainR}
The detailed results of classification accuracies achieved by baseline methods and FBCNet along with the statistical significance are provided in Table \ref{Stab:mainR}. Moreover, for the purpose of reproducibility and future comparisons, the single subject MI classification accuracies for all analyses are presented in Table \ref{tab:mainRBCCI}, \ref{tab:mainRkorea}, and \ref{tab:strokemainRes}.

\begin{table*}[htbp]
\sisetup{detect-all=true,detect-inline-weight=math}
\centering
  \begin{threeparttable}
  \caption{Subject-specific classification accuracy: All analyses (mean$\pm$std)}
    \begin{tabular}{lcllll}
    \toprule
    Dataset & Analysis & \multicolumn{1}{c}{FBCSP-SVM$^+$} & \multicolumn{1}{c}{Deep Convnet$^\ast$} & \multicolumn{1}{c}{EEGNet-8,2$^\times$} & \multicolumn{1}{c}{FBCNet$^\dagger$} \\
    \midrule
    BCIC-IV-2A Data & CV    & 75.89$\pm$13.87 & 72.20$\pm$12.12$^{\dagger}$ & 73.13$\pm$8.52 & \textbf{79.03$\pm$13.17$^{\ast}$} \\
    OpenBMI Data & CV    & 64.61$\pm$17.07$^{\times\dagger}$ & 68.33$\pm$12.34$^{\times\dagger}$ & 70.89$\pm$13.01$^{+\ast\dagger}$ & \textbf{74.70$\pm$13.95$^{+\ast\times}$} \\
    Stroke Data: A & CV    & 71.37$\pm$14.52$^{\dagger}$ & 68.81$\pm$12.02$^{\dagger}$ & 69.15$\pm$12.94$^{\dagger}$ & \textbf{79.16$\pm$14.06$^{+\ast\times}$} \\
    Stroke Data: B & CV    & 74.14$\pm$14.35$^{\dagger}$ & 71.11$\pm$8.79$^{\times\dagger}$ & 73.48$\pm$8.05$^{\ast\dagger}$ & \textbf{81.11$\pm$10.44$^{+\ast\times}$} \\
    BCIC-IV-2A Data & HO    & 68.06$\pm$14.11$^{\dagger}$ & 72.22$\pm$14.35 & 73.15$\pm$9.29 & \textbf{76.20$\pm$11.97$^{+}$} \\
    Korea Uni. Data & HO    & 60.36$\pm$14.97$^{\dagger}$ & 60.77$\pm$11.42$^{\times\dagger}$ & 63.63$\pm$11.08$^{\ast}$ & \textbf{67.19$\pm$14.38$^{+\ast}$} \\
    \bottomrule
    \end{tabular}%
  \label{Stab:mainR}%
    \begin{tablenotes}
      \footnotesize
      \item  The best performing method for each analysis is highlighted in boldface. To indicate statistically significant difference (p $<$ 0.05) between any two methods we have marked every method (column) with a call-sign ($^+$, $^\ast$, $^\times$, $^\dagger$, $^\ddagger$). The presence of any method's call-sign in any other method's column represents a statistically significant difference in classification accuracies between those two methods. CV: 10-fold cross validation, HO: Hold out analysis with separate test session.
    \end{tablenotes}
  \end{threeparttable}
\end{table*}

\begin{table}[htbp]
  \centering
  \caption{Classification accuracies for each subject in BCIC-IV-2A Dataset. }
    \footnotesize
    \begin{tabular}{ccccccccc}
    \toprule
    \multirow{2}[4]{8mm}{Subject No.} & \multicolumn{4}{c}{10-fold cross validation} & \multicolumn{4}{c}{Hold Out } \\
\cmidrule(lr){2-5} \cmidrule(lr){6-9}         & FBCSP-SVM & Deep Convnet & EEGNet-8,2 & FBCNet & FBCSP-SVM & Deep Convnet & EEGNet-8,2 & FBCNet \\
    \midrule
    1     & 85.31 & 71.03 & 72.86 & 85.76 & 77.78 & 78.13 & 79.51 & 85.42 \\
    2     & 64.51 & 52.05 & 56.25 & 61.07 & 55.56 & 45.14 & 61.11 & 60.42 \\
    3     & 90.00 & 82.41 & 83.39 & 94.51 & 79.51 & 85.42 & 88.54 & 90.63 \\
    4     & 64.02 & 58.93 & 67.54 & 68.84 & 63.19 & 67.01 & 71.53 & 76.39 \\
    5     & 73.66 & 73.57 & 76.38 & 82.54 & 53.47 & 77.43 & 71.18 & 74.31 \\
    6     & 52.72 & 62.50 & 67.05 & 58.71 & 46.88 & 53.13 & 59.03 & 53.82 \\
    7     & 92.10 & 79.33 & 73.53 & 93.08 & 86.81 & 86.46 & 71.53 & 84.38 \\
    8     & 88.62 & 82.41 & 80.27 & 86.21 & 81.25 & 78.13 & 80.56 & 79.51 \\
    9     & 72.10 & 87.59 & 80.94 & 80.54 & 68.06 & 79.17 & 75.35 & 80.90 \\
    \midrule
    Avg   & 75.89 & 72.20 & 73.13 & \textbf{79.03} & 68.06 & 72.22 & 73.15 & \textbf{76.20} \\
    Std   & 13.87 & 12.12 & 8.52  & 13.17 & 14.11 & 14.35 & 9.29  & 11.97 \\
    \bottomrule
    \end{tabular}%
  \label{tab:mainRBCCI}%
\end{table}%

\begin{table}[htbp]
  \centering
  \caption{Classification accuracies for each subject in OpenBMI Dataset. }
    \footnotesize
    \begin{tabular}{ccccccccc}
    \toprule
    \multirow{2}[4]{8mm}{Subject No.} & \multicolumn{4}{c}{10-fold cross validation} & \multicolumn{4}{c}{Hold out } \\
\cmidrule(lr){2-5} \cmidrule(lr){6-9}         & FBCSP-SVM & Deep Convnet & EEGNet-8,2 & FBCNet & FBCSP-SVM & Deep Convnet & EEGNet-8,2 & FBCNet \\
    \midrule
    1     & 64.00 & 53.00 & 56.00 & 72.50 & 69.50 & 56.50 & 56.00 & 69.50 \\
    2     & 99.00 & 73.50 & 73.00 & 97.50 & 63.00 & 62.50 & 70.50 & 73.50 \\
    3     & 91.00 & 65.50 & 78.50 & 94.00 & 92.00 & 65.00 & 75.00 & 92.50 \\
    4     & 52.50 & 79.50 & 78.00 & 78.00 & 46.50 & 65.50 & 72.50 & 75.50 \\
    5     & 90.50 & 78.50 & 79.00 & 81.50 & 69.00 & 54.00 & 61.50 & 67.00 \\
    6     & 68.00 & 89.00 & 89.00 & 89.00 & 68.50 & 92.00 & 91.50 & 87.00 \\
    7     & 63.00 & 78.00 & 84.00 & 73.50 & 56.50 & 49.50 & 61.50 & 51.00 \\
    8     & 56.50 & 81.50 & 83.00 & 76.00 & 51.50 & 63.00 & 66.50 & 68.00 \\
    9     & 52.00 & 51.50 & 49.00 & 71.00 & 48.00 & 53.00 & 49.50 & 64.00 \\
    10    & 51.00 & 55.50 & 56.50 & 64.00 & 52.00 & 53.50 & 53.00 & 62.00 \\
    11    & 53.50 & 57.00 & 65.50 & 60.00 & 52.00 & 56.50 & 50.00 & 50.50 \\
    12    & 62.50 & 81.50 & 81.50 & 79.00 & 52.00 & 58.50 & 62.50 & 57.50 \\
    13    & 67.00 & 71.00 & 68.50 & 66.00 & 47.50 & 54.50 & 55.50 & 50.50 \\
    14    & 48.00 & 63.00 & 64.00 & 58.50 & 50.50 & 57.00 & 58.50 & 53.00 \\
    15    & 52.50 & 91.50 & 97.00 & 89.50 & 48.00 & 68.50 & 67.00 & 61.50 \\
    16    & 59.00 & 62.00 & 71.50 & 73.00 & 52.00 & 76.00 & 86.00 & 77.00 \\
    17    & 54.50 & 59.50 & 64.00 & 67.50 & 49.50 & 52.50 & 56.00 & 60.50 \\
    18    & 89.00 & 49.50 & 55.50 & 92.50 & 76.50 & 45.00 & 57.00 & 81.00 \\
    19    & 80.00 & 61.00 & 69.00 & 77.00 & 64.50 & 53.00 & 66.50 & 57.50 \\
    20    & 41.50 & 62.00 & 60.50 & 60.00 & 43.50 & 56.50 & 60.00 & 60.50 \\
    21    & 93.50 & 56.00 & 69.50 & 97.00 & 92.00 & 52.00 & 65.00 & 99.00 \\
    22    & 80.00 & 58.50 & 52.50 & 82.00 & 57.50 & 55.50 & 53.00 & 72.00 \\
    23    & 54.00 & 55.00 & 46.50 & 57.00 & 61.50 & 46.50 & 48.00 & 60.50 \\
    24    & 49.00 & 50.00 & 53.50 & 49.50 & 56.00 & 46.50 & 50.50 & 52.50 \\
    25    & 50.00 & 67.00 & 70.50 & 58.00 & 53.50 & 85.00 & 81.00 & 54.50 \\
    26    & 50.00 & 79.50 & 78.50 & 70.00 & 52.50 & 72.50 & 77.00 & 71.00 \\
    27    & 50.00 & 77.50 & 80.50 & 69.50 & 54.00 & 73.50 & 76.50 & 66.00 \\
    28    & 97.00 & 95.50 & 95.00 & 93.00 & 90.50 & 60.50 & 67.50 & 72.50 \\
    29    & 89.50 & 83.50 & 81.50 & 90.00 & 97.00 & 93.50 & 66.50 & 95.00 \\
    30    & 73.00 & 72.50 & 74.00 & 78.00 & 60.50 & 58.00 & 59.00 & 61.50 \\
    31    & 57.00 & 83.00 & 84.00 & 77.00 & 51.50 & 71.50 & 71.00 & 56.50 \\
    32    & 72.50 & 58.50 & 57.50 & 79.50 & 69.50 & 54.50 & 56.00 & 73.50 \\
    33    & 87.50 & 75.50 & 75.50 & 96.00 & 72.00 & 65.50 & 61.50 & 90.50 \\
    34    & 49.00 & 63.00 & 64.00 & 56.00 & 45.50 & 58.50 & 58.00 & 54.00 \\
    35    & 63.00 & 100.00 & 100.00 & 100.00 & 54.50 & 87.00 & 90.00 & 86.50 \\
    36    & 97.50 & 73.00 & 95.50 & 98.00 & 97.50 & 77.00 & 90.00 & 99.00 \\
    37    & 89.50 & 43.00 & 47.00 & 91.00 & 89.00 & 50.00 & 49.50 & 87.50 \\
    38    & 57.50 & 66.50 & 67.50 & 63.50 & 50.50 & 52.00 & 72.00 & 49.50 \\
    39    & 57.00 & 68.50 & 69.50 & 67.50 & 49.50 & 53.00 & 62.00 & 66.00 \\
    40    & 46.50 & 61.50 & 62.00 & 60.50 & 51.00 & 70.00 & 64.00 & 61.00 \\
    41    & 57.00 & 59.00 & 57.00 & 60.00 & 49.00 & 48.50 & 53.50 & 48.50 \\
    42    & 50.50 & 66.50 & 69.00 & 59.00 & 52.00 & 68.00 & 68.00 & 67.50 \\
    43    & 80.50 & 81.00 & 86.50 & 80.00 & 68.50 & 51.50 & 50.50 & 50.00 \\
    44    & 92.50 & 75.50 & 83.00 & 97.50 & 94.50 & 59.50 & 78.00 & 99.00 \\
    45    & 84.50 & 63.00 & 65.00 & 87.50 & 70.50 & 50.00 & 55.50 & 78.50 \\
    46    & 47.50 & 69.00 & 70.50 & 70.00 & 54.00 & 66.00 & 58.00 & 72.50 \\
    47    & 49.50 & 66.00 & 73.00 & 72.50 & 49.00 & 69.00 & 71.00 & 71.00 \\
    48    & 48.00 & 85.00 & 83.50 & 81.50 & 45.00 & 55.50 & 59.50 & 57.00 \\
    49    & 58.00 & 63.00 & 66.00 & 66.00 & 49.00 & 50.50 & 49.00 & 50.50 \\
    50    & 47.50 & 57.00 & 55.50 & 57.00 & 55.00 & 54.50 & 53.00 & 51.00 \\
    51    & 53.00 & 70.00 & 74.00 & 67.50 & 51.50 & 59.50 & 61.50 & 63.00 \\
    52    & 63.00 & 63.00 & 75.50 & 75.00 & 61.50 & 67.00 & 68.00 & 62.50 \\
    53    & 45.00 & 61.50 & 66.50 & 55.00 & 52.00 & 58.50 & 63.00 & 54.00 \\
    54    & 54.50 & 59.50 & 55.50 & 52.50 & 50.00 & 48.50 & 53.00 & 55.50 \\
    \midrule
    Avg   & 64.61 & 68.33 & 70.89 & \textbf{74.70} & 60.36 & 60.77 & 63.63 & \textbf{67.19} \\
    Std   & 17.07 & 12.34 & 13.01 & 13.95 & 14.97 & 11.42 & 11.08 & 14.38 \\
    \bottomrule
    \end{tabular}%
  \label{tab:mainRkorea}%
\end{table}%

\begin{table}[htbp]
  \centering
  \caption{Classification accuracies for stroke patients from both stroke datasets. }
    \footnotesize
    \begin{tabular}{ccccccccc}
    \toprule
    \multirow{2}[4]{*}{Subject No.} & \multicolumn{4}{c}{Stroke Data: A} & \multicolumn{4}{c}{Stroke Data: B} \\
\cmidrule(lr){2-5} \cmidrule(lr){6-9}         & FBCSP-SVM & Deep Convnet & EEGNet-8,2 & FBCNet & FBCSP-SVM & Deep Convnet & EEGNet-8,2 & FBCNet \\
    \midrule
    1     & 74.23 & 68.65 & 73.82 & 73.64 & 73.75 & 66.88 & 68.66 & 70.80 \\
    2     & 51.70 & 49.11 & 45.54 & 49.29 & 78.54 & 65.79 & 71.42 & 76.58 \\
    3     & 96.79 & 95.36 & 94.73 & 97.32 & 59.55 & 68.93 & 71.79 & 77.86 \\
    4     & 49.55 & 47.95 & 43.75 & 50.54 & 46.21 & 61.42 & 65.79 & 64.92 \\
    5     & 77.41 & 77.68 & 78.48 & 91.61 & 84.79 & 73.92 & 80.29 & 88.58 \\
    6     & 96.88 & 80.64 & 80.73 & 96.16 & 82.14 & 90.98 & 88.48 & 88.39 \\
    7     & 67.32 & 73.04 & 70.80 & 74.29 & 82.29 & 77.79 & 79.67 & 82.13 \\
    8     & 90.45 & 83.13 & 87.59 & 93.48 & 55.08 & 50.00 & 56.46 & 59.04 \\
    9     & 77.14 & 56.16 & 58.57 & 77.32 & 82.95 & 77.23 & 81.61 & 86.70 \\
    10    & 78.75 & 71.79 & 79.55 & 90.27 & 53.04 & 72.83 & 65.79 & 66.92 \\
    11    & 86.43 & 76.70 & 70.18 & 96.79 & 73.66 & 69.02 & 69.55 & 79.20 \\
    12    & 56.07 & 60.63 & 59.29 & 69.38 & 86.08 & 67.79 & 72.25 & 86.08 \\
    13    & 53.52 & 58.17 & 48.09 & 58.17 & 91.16 & 78.66 & 76.70 & 92.50 \\
    14    & 61.95 & 60.21 & 64.58 & 76.63 & 53.92 & 73.50 & 74.67 & 74.13 \\
    15    & 96.13 & 81.42 & 83.92 & 97.42 & 80.63 & 65.45 & 68.93 & 88.57 \\
    16    & 62.75 & 60.13 & 61.01 & 69.84 & 65.79 & 65.08 & 70.71 & 65.17 \\
    17    & 43.76 & 57.05 & 51.12 & 50.28 & 53.83 & 72.75 & 70.79 & 67.75 \\
    18    & 64.14 & 67.85 & 69.36 & 77.70 & 73.21 & 61.88 & 67.59 & 83.30 \\
    19    & 79.84 & 75.26 & 77.48 & 82.25 & 90.00 & 70.00 & 73.39 & 92.50 \\
    20    & 60.88 & 78.38 & 77.04 & 71.92 & 72.05 & 67.86 & 65.89 & 84.29 \\
    21    & 71.14 & 63.15 & 59.80 & 79.61 & 57.14 & 62.32 & 67.95 & 68.04 \\
    22    & 77.46 & 86.63 & 84.75 & 93.08 & 86.67 & 72.75 & 77.17 & 79.08 \\
    23    & 73.54 & 67.21 & 72.33 & 87.04 & 78.42 & 68.92 & 72.21 & 84.08 \\
    24    & 62.23 & 80.18 & 74.29 & 82.68 & 81.34 & 72.23 & 81.88 & 92.41 \\
    25    & 71.46 & 72.18 & 81.76 & 86.57 & 53.13 & 67.08 & 70.96 & 72.75 \\
    26    & 67.68 & 62.68 & 61.61 & 75.36 & 91.96 & 78.30 & 79.91 & 94.02 \\
    27    & 91.61 & 87.78 & 87.07 & 96.25 & 83.75 & 72.77 & 67.86 & 88.66 \\
    28    & 56.88 & 53.30 & 56.34 & 66.70 & 87.95 & 71.70 & 74.11 & 92.50 \\
    29    & 79.57 & 74.48 & 69.94 & 90.50 & 66.92 & 66.92 & 72.00 & 75.17 \\
    30    & 92.95 & 79.89 & 84.52 & 93.44 & 81.52 & 67.14 & 61.43 & 80.18 \\
    31    & 81.25 & 78.04 & 78.66 & 86.88 & 97.50 & 81.13 & 82.25 & 99.33 \\
    32    & 60.98 & 61.61 & 61.52 & 70.98 & 50.08 & 58.13 & 66.25 & 69.67 \\
    33    & 60.67 & 58.38 & 54.88 & 61.00 & 75.98 & 94.46 & 94.64 & 91.52 \\
    34    & 71.84 & 47.32 & 53.02 & 78.99 & 89.64 & 85.98 & 89.20 & 94.82 \\
    35    & 67.96 & 62.29 & 67.92 & 85.46 &       &       &       &  \\
    36    & 46.07 & 56.25 & 59.20 & 60.54 &       &       &       &  \\
    37    & 81.71 & 75.47 & 75.51 & 89.65 &       &       &       &  \\
    \midrule
    Avg   & 71.37 & 68.81 & 69.15 & \textbf{79.16} & 74.14 & 71.11 & 73.48 & \textbf{81.11} \\
    Std   & 14.52 & 12.02 & 12.94 & 14.06 & 14.35 & 8.79  & 8.05  & 10.44 \\
    \bottomrule
    \end{tabular}%
  \label{tab:strokemainRes}%
\end{table}%

\FloatBarrier
\bibliographystyle{IEEEtran}
\bibliography{IEEEabrv,ref.bib}